\documentclass[aps, prl, twocolumn,groupedaddress,10pt,a4paper, showpacs]{revtex4-1}

\usepackage{natbib}
\usepackage{bm}
\usepackage{amsmath}
\usepackage{amssymb}
\usepackage{amsfonts}
\usepackage{graphicx}

\def\bra#1{\mathinner{\langle{#1}|}}
\def\ket#1{\mathinner{|{#1}\rangle}}
\def\braket#1{\mathinner{\langle{#1}\rangle}}
{\catcode`\|=\active
  \gdef\set#1{\mathinner{\lbrace\,{\mathcode`\|"8000\let|\midvert #1}\,\rbrace}}
  \gdef\Set#1{\left\{\:{\mathcode`\|"8000\let|\SetVert #1}\:\right\}}}
\def\midvert{\egroup\mid\bgroup}
\def\SetVert{\egroup\;\mid@vertical\;\bgroup}

\def\dm{\bm{\rho}_\alpha(t)} % Shorthand notation for the reduced density matrix

\begin{document}

%\title{Temporal Scaling of Entropy Growth for Dissipative Quantum Walks} 
\title{Information Dimension of Dissipative Quantum Walks} 
\author{P. Schijven}
\email{petrus.schijven@physik.uni-freiburg.de}
\author{O. M\"ulken}
\email{muelken@physik.uni-freiburg.de}

\affiliation{Physikalisches Institut, Universit\"at Freiburg, Hermann-Herder-Strasse 3, 79104 Freiburg, Germany}
\date{\today}

\begin{abstract}
We study the temporal growth of the von Neumann entropy for dissipative quantum walks on networks. 
By using a phenomenological quantum master equation, the quantum stochastic walk (QSW), we are able 
to parametrically scan the crossover from purely coherent quantum walks to purely diffusive random walks. 
In the latter limit the entropy shows a logarithmic growth, which is proportional to the information dimension 
of the random walk on the network. 
Here we present results for the von Neumann entropy based on the reduced density operator of the 
QSW. It shows a similar logarithmic growth for a wide range of parameter values and networks. As a 
consequence, we propose the logarithmic growth rate of the von Neumann entropy to be a natural extension 
of the information dimension to dissipative quantum systems.  We corroborate our results by comparing 
to numerically exact simulations. 
\end{abstract}

\maketitle

Much effort has been put into understanding dynamical properties of of excitation transfer on various networks 
\cite{Muelken2011}. Based on the continuous-time quantum walk (CTQW) \cite{Farhi1998}, one of the intriguing 
questions is if it is possible to achieve a classification of its quantum dynamics, similar to the classical 
universality classes \cite{Hughes1995}. For continuous-time random walks (CTRW), this can be done by studying the scaling 
exponents of dynamical properties such as the mean square displacement, the mean first passage time and the average 
return probability \cite{VanKampen1990, Havlin2002}. Due to the highly oscillatory nature of the CTQW however, 
many of these exponents do not translate well to the quantum regime \cite{Muelken2011}. In this letter, 
we show that this problem can be approached by studying the so-called information dimension associated with a 
CTQW on a network in a dissipative environment.

For classical random walks on networks, the information dimension is a dynamical exponent which is defined 
by the logarithmic growth of the entropy \cite{Argyrakis1987, Argyrakis1989}. Recently, much interest has been 
shown in the growth of the (entanglement) entropy in diffusive and disordered systems 
\cite{Serbyn2013, Kim2013, Bardarson2012}. Additionally, much attention has recently been 
given to understand the entropy production rate in open quantum systems \cite{Leggio2013, Deffner2011}. 
In this letter we provide a new perspective on the entropy growth of dissipative quantum walks by relating 
it to a generalized form of the information dimension. 

We first provide the formalism that allows us to describe the (reduced) dynamics of an excitation in an 
open quantum system/network.
When the environmental correlation time is small compared to the relaxation time of the
network, one can describe the dynamics (of the reduced density matrix $\bm{\rho}(t)$) on the network by a master equation in
Lindblad form \cite{BreuerOpenQS}:
\begin{equation}\label{eq:lindblad}
 \frac{d\bm{\rho}(t)}{dt} = -i \left[ \mathbf{H}_0, \bm{\rho}(t) \right] + \sum_{k,l=1}^N \lambda_{kl} \mathcal{D}[\mathbf{L}_{kl}, \bm{\rho}(t)],
\end{equation}
where $\mathbf{H}$ is the Hamiltonian of the (closed) system/network of $N$ nodes, where each node $k$ is associated 
with a basis vector $\ket{k}$ ($k=1,\dots,N$), all of which span the accessible Hilbert space for the closed system 
\cite{Muelken2011}. The rate constants satisfy $\lambda_{kl}\geq 0$ for all $k$ and $l$. The dissipator is given by
\begin{equation}\label{eq:dissipator}
 \mathcal{D}[\mathbf{L}_{kl}, \bm{\rho}(t)] = 
 \mathbf{L}_{kl}^{\phantom{\dag}} \bm{\rho}(t) \mathbf{L}_{kl}^\dag - 
 \frac{1}{2}\left\{\mathbf{L}_{kl}^\dag\mathbf{L}_{kl}^{\phantom{\dag}}, \bm{\rho}(t)\right\},
\end{equation}
with the Lindbald operators $\mathbf{L}_{kl}$. 
Depending on the strength of the coupling to the environment, different types of dynamics can occur. In the
case of weak coupling the dynamics is mostly coherent, which we model by CTQW. Following the formalism of the CTQW, 
the network's Hamiltonian is then chosen 
to be directly proportional to the connectivity matrix $\mathbf{A}$ of the network: $\mathbf{H}_0 \equiv \mathbf{A}$
\cite{Farhi1998,Muelken2011}. 
%This choice for $\mathbf{H}_0$ implies that there is a static diagonal disorder for the site energies,where the disorder strength is proportional to the degree of the corresponding node.
On the other hand, for
high temperatures and strong coupling, we assume in what follows that the environment induces incoherent hopping, 
i.e., a CTRW between the nodes of the
network after most coherent signatures are gone \cite{BreuerOpenQS}. In order to model the CTRW, we utilize
the freedom of choosing the set of Lindblad operators $\mathbf{L}_{kl}$ \cite{BreuerOpenQS}: It can be
shown that the Lindblad operator $\mathbf{L}_{mn} = \ket{m}\bra{n}$, for $m\neq n$, models an incoherent transition
from node $\ket{n}$ to node $\ket{m}$ \cite{Rodriguez2010, Schijven2012a, Schijven2012b}. 
Following F\"orster theory or Marcus' theory of electron transport, we can then estimate the 
incoherent transition rates $k_{n\to m}$ between the nodes $\ket{n}$ and $\ket{m}$ 
% based on the networks Hamiltonian
by using Fermi's golden rule, i.e., $k_{n\to m} \sim | \braket{m | \mathbf{H}_0 | n} |^2$ 
\cite{Marcus1956, Foerster1959, Nitzan2006}.
Upon setting
$\lambda_{mn} = k_{n\to m}$ we obtain the correct CTRW from the dissipator. 

We now use the Lindblad equation to construct a phenomenological model that allows us to interpolate 
between these two limiting scenarios. 
This model is also known as the Quantum Stochastic Walk (QSW) \cite{Rodriguez2010}.
The master equation then takes the following form:
\begin{align}\label{eq:qsw}
 \frac{d \dm}{dt} &= (1 - \alpha) \mathcal{L}_{\mathrm{CTQW}}[\dm]  \nonumber \\
                  &\phantom{=} + \alpha \big(\mathcal{L}_{\mathrm{CTRW}}[\dm] + \mathcal{L}_{\mathrm{deph}}[\dm]\big)
\end{align}
with $\alpha \in [0,1]$, $\mathcal{L}_{\mathrm{CTQW}}[\dm] = -i \left[ \mathbf{H}_0, \dm \right]$. 
Furthermore, $\mathcal{L}_{\mathrm{CTRW}}[\dm]$ models that part of the dissipator in Eq. \eqref{eq:dissipator} 
which describes the environmentally induced CTRW, while $\mathcal{L}_{\mathrm{deph}}[\dm]$ is that part of the 
dissipator which induces a localized dephasing process. To model the latter, we choose the (diagonal) Lindblad operators 
$\mathbf{L}_{mm} = \ket{m}\bra{m}$ and, for simplicity, we choose the dephasing rates to be 
$\lambda_{nn} \equiv \lambda = 1$.

In order to validate our model for more realistic systems, we also use the numerically exact Hierarchy
Equations of Motion approach (HEOM) to solve the reduced dynamics of the system \cite{Tanimura1989}.
In this method, one attempts to solve the Liouville-von Neumann
equation for the full system-environment Hamiltonian $\mathbf{H}_{\mathrm{tot}}$ by constructing a hierarchy of
auxilliary density matrices. 
\footnote{Here, the environment is modeled in terms of the Caldeira-Leggett model, as a 
bath of harmonic oscillators that is linearly coupled to the system \cite{Caldeira1983}.}
This hierarchy can be truncated and computed efficiently if one assumes that the
system-environment interaction is governed by a Lorentzian spectral density \cite{Tanimura1989}:
%\begin{equation}
$ J(\omega) = (\omega \Omega\Lambda)/[2(\omega^2 + \Omega^2)]$,
%\end{equation}
where $\Lambda$ is the reorganization energy of the bath and $\Omega$ its characteristic response frequency.

Due the Markovian nature of the QSW dynamics, it approaches a thermally mixed stationary state
when $t\to\infty$ \cite{BreuerOpenQS}. This is reflected by the von Neumann entropy
\begin{equation}
 S_{\mathrm{vn}}(t, \alpha) = - \mathrm{tr}\left\{\dm \ln \dm \right\},
\end{equation}
which increases from $S_{\mathrm{vn}}(0, \alpha) = 0$ to its maximal value $\lim_{t\to\infty}S_{\mathrm{vn}}(t, \alpha) = \ln N$
for all $\alpha > 0$ \cite{BreuerOpenQS}.
Here, we are interested in the question if the growth of $S_{\mathrm{vn}}(t, \alpha)$ 
can be related to certain scaling exponents associated with the dynamics on the underlying network.
%{\bf 
%since its growth rate can be interpreted as measure of the spreading of the excitation. Furthermore, it also provides insight in the way information about the excitation is lost due to energy dissipation into the environment. 
%}

In the classical limit, ($\alpha \to 1$) $S_{\mathrm{vn}}(t, \alpha)$ reduces to the classical Shannon entropy $H(t)$ of 
the environmentally induced CTRW:
\begin{equation}
 H(t) \equiv \lim_{\alpha\to 1} S_{\mathrm{vn}}(t, \alpha)  = - \sum_k p_{kj}(t) \ln[ p_{kj}(t) ],
\end{equation}
where $p_{kj}(t) = \lim_{\alpha\to 1} \braket{k | \dm | k}$, given that $\bm{\rho}_\alpha(0) = \ket{j}\bra{j}$.
For various networks, it has been numerically shown that $H(t)$ grows linearly with $\ln(t)$ after an initial/transient time $t_I$. 
The logarithmic growth rate is now defined as the CTRW variant of the information dimension, i.e., $H(t) \sim d_I \ln t$.
We pause to note
that this definition is akin to the definition of the information dimension of chaotic systems \cite{OttChaos}. 
For a discrete-time random walk, having performed $M$ steps, it is defined as $d_I = I_M / \ln M$. Here 
$I_M = - \sum_{k=1}^{S_M} P_k \ln P_k$, with $P_k$ being the probability of visiting the $k$-th site
\cite{Argyrakis1987, Argyrakis1989}. Upon taking the limit to the CTRW, this matches
our definition.

The existence of a logarithmic growth for CTRW can be
shown analytically by assuming
that the propagator for the CTRW on a generic complex network 
can be formulated as a stretched exponential \cite{Blumen1991}, 
%\begin{equation}
$ p_{kj}(t) \sim t^{-d_s/2} \exp\left( -a \xi_k^\nu \right)$,
%\end{equation}
where $\xi_k = r_k t^{-d_s/2d_f}$. Here, $r_k$ is the position of the $k$-th node relative to 
node $j$, $d_f$ the fractal dimension and $d_s$ the spectral dimension
of the network. The latter is reflected in the scaling behaviour of the probability 
to return or to remain at the origin, i.e., $p_{jj}(t) \sim t^{-d_s/2}$ \cite{Alexander1982}. 
Substituting this form into the definition of $H(t)$ results in:
\begin{equation}
  H(t) = \frac{d_s}{2} \ln t + t^{-\nu d_s/2d_f} \sum_k r_k^\nu p_{kj}(t) \approx \frac{d_s}{2} \ln t.
\end{equation}
In this case we thus obtain $d_I = d_s / 2$

\begin{figure*}
 \begin{center}
  \includegraphics[width=\textwidth]{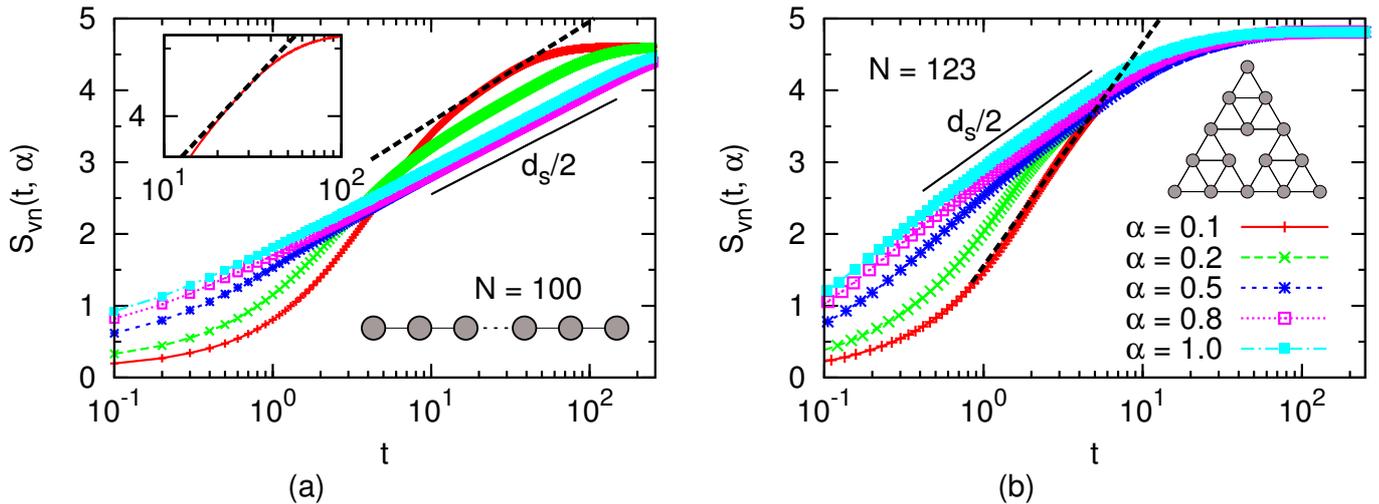}
 \end{center}
 \caption{The left panel (a) shows the von Neumann entropy $S_{\mathrm{vn}}(t, \alpha)$ for a linear chain with
 $N=100$ nodes on a log-linear scale for various values of the interpolation parameter $\alpha$. The right panel 
 (b) shows $S_{\mathrm{vn}}(t, \alpha)$ for a Sierpinski gasket of generation $g=5$ with $N=123$ nodes, 
 for the same values of $\alpha$. In both cases we observe a logarithmic growth regime for most values of $\alpha$.
 The dashed black lines illustrate the regions for $\alpha = 0.1$, where we fitted the logarithmic growth regime.
 The inset in panel (a) shows the fitted result in more detail.  
 The solid black lines illustrate the slopes of the entropy in the classical limit and
 the illustration of the Sierpinski fractal in the right panel has generation $g=3$.}
 \label{fig:entropy-line}
\end{figure*}

Since the von Neumann entropy is well-defined for all values of $\alpha$ and contains the Shannon entropy $H(t)$
as a limiting case, we now introduce the (possibly $\alpha$-dependent) information dimension $d_I(\alpha)$ for 
the QSW in a similar way as above. Provided that $S_{\mathrm{vn}}(t, \alpha)$ grows linearly with $\ln(t)$ for intermediate
times, $d_I(\alpha)$ is defined by:
\begin{equation}
 S_{\mathrm{vn}}(t, \alpha) \sim d_I(\alpha)  \ln(t).
\end{equation}
Note that for $\alpha = 0$, the system is always in a pure state, resulting in $S_{\mathrm{vn}}(t, \alpha=0) = 0$ 
for all $t$. Therefore
it is not possible to define $d_I(\alpha)$ for $\alpha = 0$ in this way.

%The von Neumann  entropy can be written in terms of the time-dependent eigenvalues $\zeta_n(t, \alpha)$ of the density operator $\dm$:
%\begin{equation}
%  S_{\mathrm{vn}}(t, \alpha) = \sum_n \zeta_n(t, \alpha) \ln \zeta_n(t, \alpha).
%\end{equation}
%The complexity of the master equation,Eq. \eqref{eq:qsw}, only allows us to determine the $\zeta_n(t, \alpha)$ analytically for very small systems. For them, however, it is difficult to extract $d_I(\alpha)$ due to large finite-size effects. 
Since an analytic solution of Eq. \eqref{eq:qsw} is only possible for a few cases and for small systems we proceed 
with a numerical analysis of two important examples. As prototypes for opposing dynamical behaviors we 
take the linear chain and the so-called Sierpinski gasket. For the former, the CTQW is more efficient, i.e.
faster than the corresponding CTRW, while for the latter the converse is true \cite{Muelken2011, Agliari2008}.
% 
% take the linear chain for which the CTQW is more efficient, i.e., faster, than the corresponding CTRW 
% and the so-called Sierpinski gasket for which the converse is true \cite{Muelken2011, Agliari2008}.
The Sierpinski gasket is a deterministic fractal structure where each new generation $g$
is determined by subdividing each triangle into three new sub-triangles 
\footnote{see the inset in Fig. \ref{fig:entropy-line}(b) for an illustration for $g=3$}
such that the number of nodes in generation $g$ is given by $N_g = 3(3^{g-1}+1)/2$. 
%{\bf
%Considering these two structures allows for a clear-cut distinction in the scaling behavior of the von Neumann entropy. 
%}

Fig. \ref{fig:entropy-line} shows the von Neumann entropy as a function of time 
(in log-lin scale) for different values of $\alpha$ for (a) a line of $N=100$ nodes 
and for (b) a Sierpinski gasket of generation $g=5$ with $N=123$ nodes. The first 
thing to notice is that at a given small (transient) time ($t<1$) the entropy in the 
CTQW limit is smaller than in the CTRW limit. In order to understand this, we consider 
a dimer, i.e. a network of only two nodes. It is straightforward to show that for times 
$t\ll1$ one obtains
%\begin{equation}
$S_{\mathrm{vn}}(t, \alpha) \approx \alpha t - \alpha t \ln(\alpha t)$.
%\end{equation}
A similar result has also been obtained for a slightly different model in Ref. \cite{Nizama2012}.

However, with increasing time, the increase in entropy becomes larger the 
smaller the value of $\alpha$. Before the entropy reaches its stationary value, 
we find the logarithmic scaling for both the line and the Sierpinski gasket. 
While the scaling region stretches over a rather long period of time for values of 
$\alpha\geq 0.2$, namely $t\in[10,100]$ for the line and $t\in[1,10]$ for the gasket, 
this region is smaller the smaller the values of $\alpha$.
For $\alpha = 0.1$ and for the finite structures considered here, the scaling regions 
become very small.
% such that that there is a large fitting error when determining $d_I(\alpha)$. 
We are certain that this issue can be resolved by computing the entropy for a 
larger chain in order to delay the approach to the equilibrium state. 
The difference in the time scales for the line and the Sierpinski gasket can be 
understood when realizing that the Sierpinski gasket of $g=5$ has a side length of $17$ 
nodes which is about one order of magnitude smaller than the line with $N=100$ nodes.
A further obvious difference between the two structures is the fact that for the line 
the entropies for different values cross at times $t\in[1,10]$ while such a crossing 
region is far less pronounced for the Sierpinski gasket. 

\begin{figure}
 \begin{center}
  \includegraphics[width=\columnwidth]{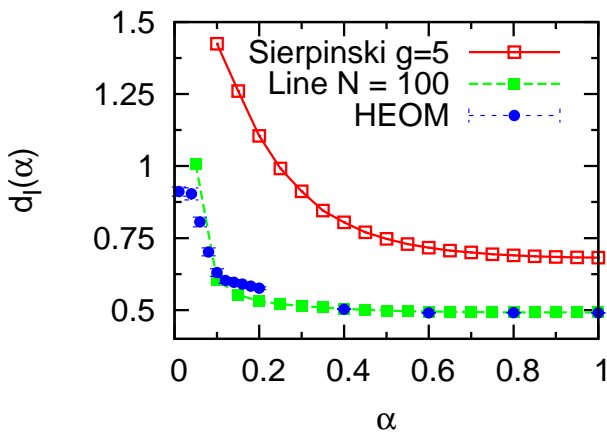}
 \end{center}
 \caption{The information dimension $d_I(\alpha)$ for both the Sierpinski gasket  of generation $g=5$ (red)
 and the linear chain with $N = 100$ nodes (green). We also show the information dimension $d_I(\Lambda)$ obtained with the HEOM,
 for a chain of $N = 9$ nodes (blue). We converted $d_I(\Lambda)$ to $d_I(\alpha)$ by choosing $\alpha = \Lambda / 500$.}
 \label{fig:information-dimension}
\end{figure}

Turning now to the scaling behavior of the entropy, we find the expected results in the CTRW limit.
In this case, the information dimension is determined by the spectral dimension. For the line the
spectral dimension is $d_s = 1$ \cite{Muelken2006}. Thus, for CTRW on a line we obtain 
$\lim_{\alpha\to 1} d_I(\alpha) = d_s/2 = 1/2$. The spectral dimension of the Sierpinsik gasket 
is also known \cite{Blumen1991, Muelken2011}: $d_s = 2 \ln 3 / \ln 5 \approx 1.365$. Thus, for 
CTRW on a Sierpinski gasket we expect $\lim_{\alpha\to 1} d_I(\alpha) \approx 0.683$, which is 
larger than the value for the line.
%; its Hamiltonian is given by \cite{Muelken2011}:
%\begin{align}
% \mathbf{H}_0^{\mathrm{line}} =& \sum_{n=2}^{N-1}\left(2\ket{n}\bra{n} -\ket{n-1}\bra{n} - \ket{n+1}\bra{n} \right) \nonumber \\
%                              & + \ket{1}\bra{1} + \ket{N}\bra{N} - \ket{2}\bra{1} - \ket{N-1}\bra{N}.
%\end{align}
%Hence, the incoherent hopping rates are given by $\lambda_{mn}=k_{n\to m}=1$ for $m\neq n$.
%The spectral dimension for a linear chain is known analytically,
%namely $d_s = 1$ {\bf citation}. This implies that for CTRW $\lim_{\alpha\to 1} d_I(\alpha) = d_s/2 = 1/2$. 

When $\alpha\to 0^+$, the QSW approaches the CTQW. Recent results for discrete-time quantum walks 
on periodic one-dimensional lattices suggest that the logarithmic growth of the von Neumann entropy 
in this limit scales with an exponent which is twice as large as the classical random walk exponent \cite{Kollar2014}. 
For our analysis this translates to $\lim_{\alpha\to 0^+} d_I(\alpha) \approx d_s = 1$.

Fig. \ref{fig:information-dimension} shows $d_I(\alpha)$ for the two structures as a function of $\alpha$. 
We have extracted the value of $d_I(\alpha)$ from the curves in Fig. \ref{fig:entropy-line} by a linear fit 
in the scaling regions. For both structures we find that $d_I(\alpha)$ increases with decreasing $\alpha$ from 
$1$ to $0.1$. For the line the values of $d_I(\alpha)$ remain close to the CTRW value $1/2$ and only start to 
increase slowly for $\alpha\leq 0.4$, e.g., for $\alpha=0.1$ we find $d_I(\alpha=0.1)\approx 0.6$. 
For smaller values of $\alpha$ there is a steep increase in $d_I(\alpha=0.1)$ up to values $d_I(\alpha=0.05) \approx 1$, 
which is twice as large as the CTRW value. The Sierpinski gasket on the other hand shows a more continuous increase 
of $d_I(\alpha)$ with decreasing $\alpha$, e.g., for $\alpha=0.1$ we find $d_I(\alpha=0.1)\approx 1.4$, 
which also is about twice as large as the CTRW value. The fact that for both
the line and the Sierpinski gasket, the information dimension is approximately twice as large in the CTQW regime as
in CTRW regime, might indicate that this is a more universal feature which certainly needs further investigation, 
see also Ref. \cite{Kollar2014} for the discrete-time quantum walk. 

\begin{figure}
 \begin{center}
  \includegraphics[width=\columnwidth]{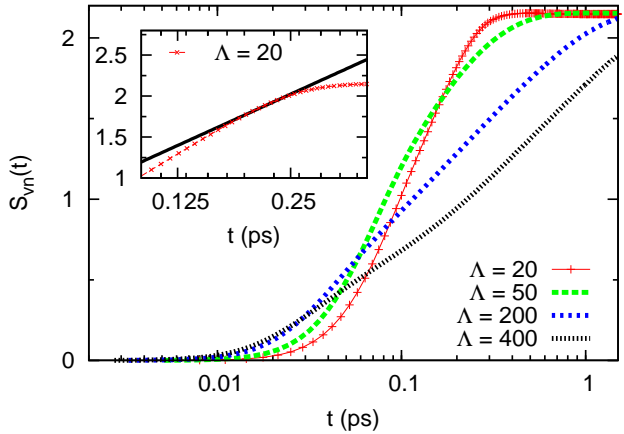}
 \end{center}
 \caption{Log-linear plot showing the von Neumann entropy $S_{\mathrm{vn}}(t)$ obtained with the 
 HEOM for different values of the reorganization energy $\Lambda$ (in units of $\mathrm{cm}^{-1}$). 
 The inset shows the area for $\Lambda = 20$, where we fitted the logarithmic growth regime. }
 \label{fig:entropy-heom}
\end{figure}

In order to corroborate our findings from our phenomenological model, we also compute the
von Neumann entropy $S_{\mathrm{vn}}(t)$ for a linear chain by using the HEOM.
% numerically exact Hierarchy Equations of Motion approach (HEOM). 
% The hierarchy can be truncated and computed efficiently when one assumes that the
% system-environment interaction is given by a Lorenzian spectral density \cite{Tanimura1989}: 
% $J(\omega) = \Lambda\Omega \omega/[2(\Omega^2 + \omega^2)]$,
% where $\Lambda$ is the reorganization energy of the bath and $\Omega$ its characteristic response frequency. 
To solve the HEOM we use the
 program PHI \cite{Strumpfer2012, PHIPackage}. Due to limited computational
resources we take a chain of $N=9$ nodes and truncate the hierarchy at a depth $L=14$,
corresponding to 497420 auxilliary density matrices. 
In particular, we have $\mathbf{H}_0 = \gamma \mathbf{A}^\mathrm{line}$, and choose
$\gamma = 50\, \mathrm{cm}^{-1}$ and $\Omega = 35\,\mathrm{ps}^{-1}$.
The temperature is chosen to be 300K in order to avoid adding low-temperature Matsubara correction terms to 
the hierarchy \cite{Ishizaki2005, Chen2011, Strumpfer2012}.
To obtain classical hopping dynamics for
large couplings to the environment we assume a broad spectral density, in accordance with the classical
theory of electron transport \cite{Nitzan2006, Marcus1956}. 

In Fig. \ref{fig:entropy-heom} we show the entropy $S_{\mathrm{vn}}(t)$ computed with the HEOM for various
values of the reorganization energy $\Lambda$, playing a similar role as the interpolation parameter
$\alpha$ for the QSW. Comparing to Fig. \ref{fig:entropy-line}(a), we observe good
overall agreement, including the crossing point for small $t$, even though the approach toward the scaling regime 
looks different. This is to be expected
since the HEOM allows for a more precise modelling of the dynamics at early times, before most of the
coherences have died out. Furthermore, having a small chain introduces finite-size effects that lead to 
a different form of $S_{\mathrm{vn}}(t)$. Indeed, the inset of Fig. \ref{fig:entropy-heom}, showing the linear fit
for $\Lambda = 20 \, \mathrm{cm}^{-1}$, has a very small linear regime, making the extraction of the information
dimension difficult.
Nonetheless, we see in Fig. \ref{fig:information-dimension} that the information dimension $d_I(\Lambda)$ (blue) has a 
similar shape as the curve for the QSW. For a qualitative comparison to our QSW results, we have 
converted $d_I(\Lambda)$ into $d_I(\alpha)$ by assuming that $\alpha = \Lambda / 500$, since 
at $\Lambda = 500\, \mathrm{cm}^{-1}$ the dynamics is purely classical, with $d_I(500) = 0.5$. 
As mentioned before, due to the small size of the chain it was difficult to extract $d_I(\Lambda)$ for 
small $\Lambda$. This is why for small $\alpha$, the curve in Fig. \ref{fig:information-dimension} does not 
reach the same value of the information dimension as for the QSW.
% but instead converges to $d_I \approx 0.92$.
% 
% 
% Nonetheless, we see in the inset of Fig. \ref{fig:entropy-heom} that
% the information dimension $d_I(\Lambda)$ has a similar shape as the corresponding curve in Fig. \ref{fig:information-dimension}:
% in the quantum regime (small $\Lambda$) it approaches the value $d_I \approx 0.92$, while for large $\Lambda$
% it approaches the classical value $d_I \approx 0.5$.

To conclude, we have seen that the information dimension $d_I(\alpha)$,
obtained from the QSW model,
is a very useful quantity for classifying the quantum dynamics on networks in
dissipative environments, in contrast to earlier methods such as the average return probability. 
% Furthermore,
% we have seen that it is robust against the particular implementation of the dissipative process, but still captures
% the essentail details of the underlying network structure. 
Additionally, the curves for $d_I(\alpha)$ can give insight into the robustness of the quantum 
dynamics against environmental noise. By comparing to numerically exact computations, we have shown
that this result is not sensitive to the particular implementation of the dissipative process, but
that it still captures the essential details of the underlying network structure.

% We gratefully acknowledge financial support from the Deutsche Forschungsgemeinschaft (DFG grant MU2925/1-1) and
% support from the DAAD. Furthermore, we would like to thank A. Anichenko for useful discussions.
We gratefully acknowledge financial support from the Deutsche Forschungsgemenschaft (DFG grant No. MU2925/1-1)
and support from the Deutscher Akademi\-scher Austauschdienst (DAAD grant No. 56266206 and project No. 40018).
Furthermore, we would like to thank A. Anischenko and A. Blumen for useful discussions. 

\end{document}